\documentclass[%
 aip,
 amsmath,amssymb,
reprint,%
]{revtex4-1}

\usepackage{graphicx}
\usepackage{dcolumn}
\usepackage{bm}

\usepackage[utf8]{inputenc}
\usepackage[T1]{fontenc}
\usepackage{mathptmx}
\usepackage{etoolbox}

\makeatletter
\def\@email#1#2{%
 \endgroup
 \patchcmd{\titleblock@produce}
  {\frontmatter@RRAPformat}
  {\frontmatter@RRAPformat{\produce@RRAP{*#1\href{mailto:#2}{#2}}}\frontmatter@RRAPformat}
  {}{}
}%
\makeatother

\begin{document}
\preprint{AIP/123-QED}

\title{Variational embedding of protein folding simulations using gaussian mixture variational autoencoders}

\author{Mahdi Ghorbani}
\email{ghorbani.mahdi73@gmail.com.}
\affiliation{ 
Laboratory of Computational Biology, National, Heart, Lung and Blood Institute, National Institutes of Health, Bethesda, Maryland 20824, USA.
}%
 \affiliation{Department of Chemical and Biomolecular Engineering, University of Maryland, College Park, MD 20742, USA}
 
\author{Samarjeet Prasad}%
\affiliation{ 
Laboratory of Computational Biology, National, Heart, Lung and Blood Institute, National Institutes of Health, Bethesda, Maryland 20824, USA.
}%

\author{Jeffery B. Klauda}
\affiliation{%
Department of Chemical and Biomolecular Engineering, University of Maryland, College Park, MD 20742, USA}%
\author{Bernard R. Brooks}%
\affiliation{ 
Laboratory of Computational Biology, National, Heart, Lung and Blood Institute, National Institutes of Health, Bethesda, Maryland 20824, USA.
}%

\date{\today}

\maketitle


Conformational sampling of biomolecules using molecular dynamics simulations often produces large amount of high dimensional data that makes it difficult to interpret using conventional analysis techniques. Dimensionality reduction methods are thus required to extract useful and relevant information. Here we devise a machine learning method, Gaussian mixture variational autoencoder (GMVAE) that can simultaneously perform dimensionality reduction and clustering of biomolecular conformations in an unsupervised way. We show that GMVAE can learn a reduced representation of the free energy landscape of protein folding with highly separated clusters that correspond to the metastable states during folding. Since GMVAE uses a mixture of Gaussians as the prior, it can directly acknowledge the multi-basin nature of protein folding free-energy landscape. To make the model end-to-end differentialble, we use a Gumbel-softmax distribution. We test the model on three long-timescale protein folding trajectories and show that GMVAE embedding resembles the folding funnel with folded states down the funnel and unfolded states outer in the funnel path.  Additionally, we show that the latent space of GMVAE can be used for kinetic analysis and Markov state models built on this embedding produce folding and unfolding timescales that are in close agreement with other rigorous dynamical embeddings such as time independent component analysis (TICA).

\section{Introduction}

In recent years, computer simulations of biomolecular systems have gained huge attention due to advances in theoretical methods, algorithms and computer hardware to efficiently explore processes in atomic scale using molecular dynamics (MD) simulations.\cite{hospital2015molecular} In a MD simulation, one integrates the Newton's equations of motion where the forces between atoms in the system are described by force field parameters. Exploration of the high dimensional space typically requires long timescale simulations or the use of some enhanced sampling techniques.\cite{yang2019enhanced, bernardi2015enhanced} These simulations usually generate a large amount of high dimensional data making analyzing the important features of protein folding such as free energy landscape (FEL) and identifying metastable states a challenging task.\cite{glielmo2021unsupervised} Therefore, dimensionality reduction techniques are often used to describe the processes such as folding and conformational transitions of proteins.\cite{lemke2019encodermap} 

The ideal FEL should consist of heavily clustered datapoints, where each cluster is positioned in a local free energy minimum and corresponds to long-lived metastable states separated by kinetic bottlenecks (i.e. free energy barriers).\cite{hegger2007complex} This ideal FEL is the cornerstone of many kinetic models that describe the dynamics of the system using for example Markov state models (MSM).\cite{chodera2006long,chodera2007automatic,chodera2014markov} Traditional methods to capture FEL, rely on identifying the relevant collective variables (CVs) that are well-suited to describe the physical processes or to distinguish different states. However, finding the right collective variables for the system of interest requires a physical/chemical intuition about the process of interest.\cite{dobson2003protein,onuchic2004theory} This makes it necessary to define a low-dimensional representation of the system that can capture the essential degrees of freedom of the CVs of the system of interest. There are various methods for dimensionality reduction and finding optimal representation of complex FEL such that employ linear (PCA\cite{abdi2010principal} and TICA\cite{schwantes2013improvements,perez2013identification}) or non-linear (Isomap\cite{balasubramanian2002isomap}, sketch map\cite{ceriotti2011simplifying}, diffusion map\cite{nadler2005diffusion, nadler2006diffusion}).

Machine learning (ML) has recently emerged as a powerful alternative tool for learning informative representations and in particular variational auotencoders (VAEs) have shown great potential for unsupervised representation learning \cite{kingma2013auto}. An autoencoder has two parts: encoder and decoder. The encoder network reduces the input data to a low-dimensional latent space and the decoder maps the latent representation back to the original data. In the VAE framework, a regularization is added to the model by forcing the latent space to be similar to a pre-defined probability distribution (e.g Gaussian) which is called a prior. VAEs have been recently used for CV discovery in MD simulations \cite{ chen2018molecular,schoberl2019predictive,chen2018collective}, enhanced sampling \cite{ribeiro2018reweighted,bonati2019neural} and dimensionality reduction methods \cite{bhowmik2018deep, varolgunecs2020interpretable}. Noe and coworkers proposed a time-lagged autoencoder (TAE) that can find the low-dimensional embedding for high dimensional data while capturing the slow dynamics of the underlying processes \cite{wehmeyer2018time}. Although Ferguson et al.\cite{chen2019capabilities} showed that TAE is limited in finding the optimal embedding for the dynamical system and in general it finds a mixture of slow and maximum variance modes. 

In a simple VAE, the prior is a simple standard distribution, which can lead to over-regularization of the posterior distribution and result in posterior collapse.\cite{guo2020variational} This makes the output of the decoder almost independent of the latent embedding and can result in poor reconstruction and highly overlapping clusters in the latent space \cite{bhowmik2018deep}. On the other hand, a Gaussian prior is limited since the learnt representation can only be unimodal and cannot capture multimodal nature of data such as protein folding simulation where there exist multiple metastable states during the folding process.\cite{dill2008protein} In this work we employ a gaussian mixture variational autoencoder (GMVAE) that directly acknowledges the multimodal nature of protein folding simulations and can construct the ideal multi-basin FEL. This is achieved by modeling the latent space as a mixture of Gaussians by using a categorical variable that identifies which mode each data point comes from. Our model is different from the GMVAE model proposed by Varolg\"{u}ne\c{s} et al.\cite{varolgunecs2020interpretable} in that instead of learning the stochastic layer to learn the cluster assignment probabilities is replaced with a deterministic layer using Gumbel-softmax distribution which makes the model end-to-end differentiable and leads to better performance.\cite{figueroa2017simple,jang2017categorical} Moreover, we make use of convolutional layers since they use sliding filter maps to capture local patterns in the data independent of its position. It is worth noting that, our GMVAE model is different from a simple Gaussian mixture model (GMM). In a GMM, the parameters of the model are optimized iteratively through expectation-maximization algorithm.\cite{dempster1977maximum} GMM has been used to cluster the FEL of proteins. Delemotte et al. used GMM to construct and cluster the FEL of binding $Ca^{2+}$ to Calmodulin and found novel pathway involving salt bridge breakage and formation.\cite{westerlund2019inflecs} However, GMM requires the use of a few handcrafted features and a high number of collective variables can lead to over-fitting the model. On the other hand, since the GMVAE model is trained by gradient descent and is a deep learning architecture, it does not suffer from the same shortcomings of GMM.

The GMVAE model simultaneously performs dimensionality reduction and clustering.\cite{dilokthanakul2016deep} The number of clusters is a hyperparameter of the model that needs to be pre-defined. Varolg\"{u}ne\c{s} et al.\cite{varolgunecs2020interpretable} used a thresholding scheme to pick the clusters that have class probabilities that are more than a cutoff value (e.g 0.95). However, we found that the number of clusters could largely vary depending on the cutoff value chosen for larger proteins studied here. To make a hard-cluster assignment, we use a k-nearest neighbors method to assign each datapoint to the most likely cluster in its neighborhood. The features in our model are the normalized distance map between $C_{\alpha}$ atoms of the protein. We test our model on three long-timescale protein folding simulations taken from DE Shaw group \cite{lindorff2011fast}. These include Trp-cage (208 $\mu s$), BBA (325 $\mu s$) and Villin (125 $\mu s$). We show that the model can learn the funnel-shaped landscape of protein folding and cluster the conformational space with high accuracy that correspond to different structural features of protein. The funnelness of the folding landscape can be viewed from the conformational entropy point of view where the folded state has a narrower distribution than the unfolded state. Furthermore, we show that despite the fact that the GMVAE embedding does not make use of any dynamical information, it is able to describe the kinetics of protein folding and the folding and unfolding timescales obtained by making a Markov model on this embedding are in close agreement with other works using a rigorous dynamical model to describe the kinetics.

\section{Methods}

Variational inference methods convert an intractable inference problem into an optimization one.
While the classical variational methods are limited to conjugate priors and likelihood, VAEs allow the use of arbitrary function approximators (i.e. neural networks) as the conditional posterior \cite{kingma2013auto}.
Using the reparameterization trick, standard backpropagation can be to optimize the variational objective. 

VAEs can be approached from two different perspectives: variational inference and neural networks. In the variational inference, the main idea is to learn a distribution in the latent space that truly captures the distribution of the dataset. In particular, given a dataset $x$, the goal of variational inference is to infer the latent space representation $z$ i.e. to accurately model $p(z|x)$. The Bayes theorem gives the relation between the posterior $p(z|x)$, the prior $p(z)$ and the likelihood $p(x|z)$ as:

\begin{equation}
    p(z|x)=\frac{p(x|z)p(z)}{p(x)}
\end{equation}

The denominator in this equation $p(x)$ is called the evidence which requires marginalization over all latent variables and thus is intractable. Therefore, in variational inference one seeks an approximate posterior $q_\phi(z|x)$ with learnable parameters $\phi$ and minimize the Kullback-Leibler divergence (KL) between the approximate and true posterior. The KL divergence shows the difference between two probability distributions and is defined as:

\begin{equation}
    D_{KL}(q_\phi(z|x)||p(z|x)) = \mathbb{E}_{q} \log(\frac{q_\phi(z|x)}{p(z|x)})
\end{equation}

by re-writing this equation and using Bayes rule we get:

\begin{equation}
     \log p(x) =  KL(q_\phi(z|x)||p(z|x)) + \mathbb{E}_{q} \log({\frac{q_\phi(z|x)}{p(x,z)}})
\end{equation}

Due to Jensen's inequality the KL divergence is a non-negative term which makes the last term in the equation called evidence lower bound (ELBO) to act as a lower bound for the log-likelihood of the evidence.

\begin{equation}
    ELBO = \mathbb{E}_{q} \log({\frac{p(x,z)}{q_\phi(z|x)}})
\end{equation}

Therefore, we can now write equation 3 as:

\begin{equation}
 \log p(x) =  KL(q_\phi(z|x)||p(z|x)) + ELBO
\end{equation}

This has the implication that minimizing the KL divergence or maximizing the log-likelihood of evidence can done by maximizing the ELBO.

The graphical model of GMVAE is shown in figure 1A. In the generative part (decoder) of the network, a sample $z$ is drawn from the latent space distribution $p_\beta(z|y)$ of cluster $y$ which is parameterized by parameters $\beta$ using the decoder part of the neural network. This can be used to generate the conditional distribution $p_\theta(x|z)$ parameterized by another neural network $\theta$. The generative process for GMVAE can be written as 

\begin{equation}
    p_{\beta,\theta}(x,z,y) = p_\theta (x|z) p_\beta (z|y) p(y)
\end{equation}
\begin{equation}
    p_\beta (z|y) = N(z|\mu_\beta(y),\sigma^2_\beta (y))
\end{equation}
\begin{equation}
    p_\theta (z|x) = N(x|\mu_\theta(z),\sigma^2_\theta (z))
\end{equation}
\begin{equation}
    p(y) = Cat(\pi)
\end{equation}

In these equations, $\pi=1/K$ is the uniform categorical distribution where K is the number of clusters, $N()$ refers to normal distribution where $\mu_\theta$, $\mu_\beta$, $\sigma^2_\theta$, $\sigma^2_\beta$ are the means and variances learned by the neural nets parameterized by $\theta$ and $\beta$. 
Variational inference of GMVAE can be done by maximizing the ELBO which can be written as:
\begin{equation}
    ELBO = \mathbb{E}_{q} \log{\frac{p_{\beta,\theta}(x,z,y)}{q_{\phi,\psi}(z,y|x)}}
\end{equation}
The approximate posterior of the inference model $q_{\phi,\psi}(z,y|x)$ can be factorized into two distributions:
\begin{equation}
   q_{\phi,\psi}(z,y|x) = q_\phi(y|x) q_\psi (z|x,y)
\end{equation}
$q_\phi(y|x)$ gives the cluster assignment probabilities and thus $\sum_{k=1}^{K} q_\phi(y|x)=1$. $q_\psi(z|x,y)$ is a Gaussian mixture where the parameters of the each Gaussian $(\mu_\psi, \sigma^2_\psi$) are learned by the encoder part of neural network. In this model, categorical variable $y$ represents a discrete node for each categorical distribution, which cannot be backpropagated and thus is substituted with a Gumbel-softmax distribution which approximates this categorical distribution with a continuous one. This can be written as:
\begin{equation}
    y_i = \frac{e^\frac{\log(\pi_i)+g_i}{\tau}}{\sum_{j=1}^{K}e^\frac{\log(\pi_j)+g_j}{\tau}}  \hspace{4mm} {for \hspace{4mm} i=1...K}
\end{equation}
where $\tau$ is called the temperature parameter controls the smoothness of distribution where at small temperatures samples are close to one-hot encoded and at large temperatures, the distribution is more smooth. $g_i$ are the samples drawn from a Gumbel(0,1) distribution.

Using the generative and inference model the ELBO can be written as:

\begin{equation}
    ELBO = \mathbb{E}_{q} \log{\frac{p_\theta(x|z)p_\beta(z|y)p(y)}{q_\phi(y|x)q_\psi(z|x,y)}}
\end{equation}

\begin{equation}
\begin{aligned}
    ELBO = \mathbb{E}_{q} [\log{p(y)} - \log{q_\phi(y|x)} + \\ \log{\frac{p_\beta(x|z)}{q_\psi(z|x,y)}}+\log{p_\theta(x|z)}]
\end{aligned}
\end{equation}

The second term in the loss is called the cross-entropy and the last term is the mean squared error between the true and the reconstructed data. 

\subsection{Model parameters}
The network architecture is shown in figure 1B. The GMVAE model was implemented in tensorflow. Convolutional layers were applied along with pooling for their ability to recognize features in images. Exponential linear unit (Elu) activation function was used in each layer and a softmax activation was used for the cluster assignment probability. The means and variances of distributions were obtained using no activation, and softplus activation respectively. Adam was used as optimizer in all models.\cite{kingma2014adam}  We have optimized the hyperparameters of the model based on the reconstruction loss and the cross-entropy term. The chosen hyperparameters for each protein are shown in table 1. Embeddings obtained by training the GMVAE model with different hyperparameters, mainly different number of clusters are shown in supplementary information (SI).  Each model was trained for 100 epochs of training. We also tried annealing the temperature parameter starting with a high value (5) and lowering it to 0.1 during the first 40 epochs of training and then keeping it the same for the rest of training. However, we found that the model would diverge after a few epochs of training and having a fixed and small value of temperature parameter gives the best results. Since the GMVAE model gives a probabilistic cluster assignment that is the probability of each datapoint belonging to each cluster, we used a k-nearest neighbors method to compute a hard-cluster assignment using the neighborhood of each point in the embedding. For the kinetic analysis, we used pyemma package \cite{scherer_pyemma_2015} to build the transition matrix. In each case, the embedding was discretized using 500 K-means cluster points and the transition probability matrix was built by counting the number of transitions between different states at lag time $\tau$. The implied time scales are computed from the eigenvalues of the transition probability matrix:
\begin{equation}
    t_{i}(\tau) = -\frac{\tau}{\ln{|\lambda_{i}(\tau)|}}
\end{equation}
To test the Markovianity of the transition matrix the implied timescales are plotted against the lag-time and then the smallest $\tau$ is chosen such that the implied timescales have converged. A coarse-grained transition matrix is later built by assigning the K-means points to the closest GMVAE clusters yielding a coarse-grained view of dynamics. The folding and unfolding timescales are obtained from this coarse-grained picture.

\begin{figure*}
    \centering
    \includegraphics[scale=0.54]{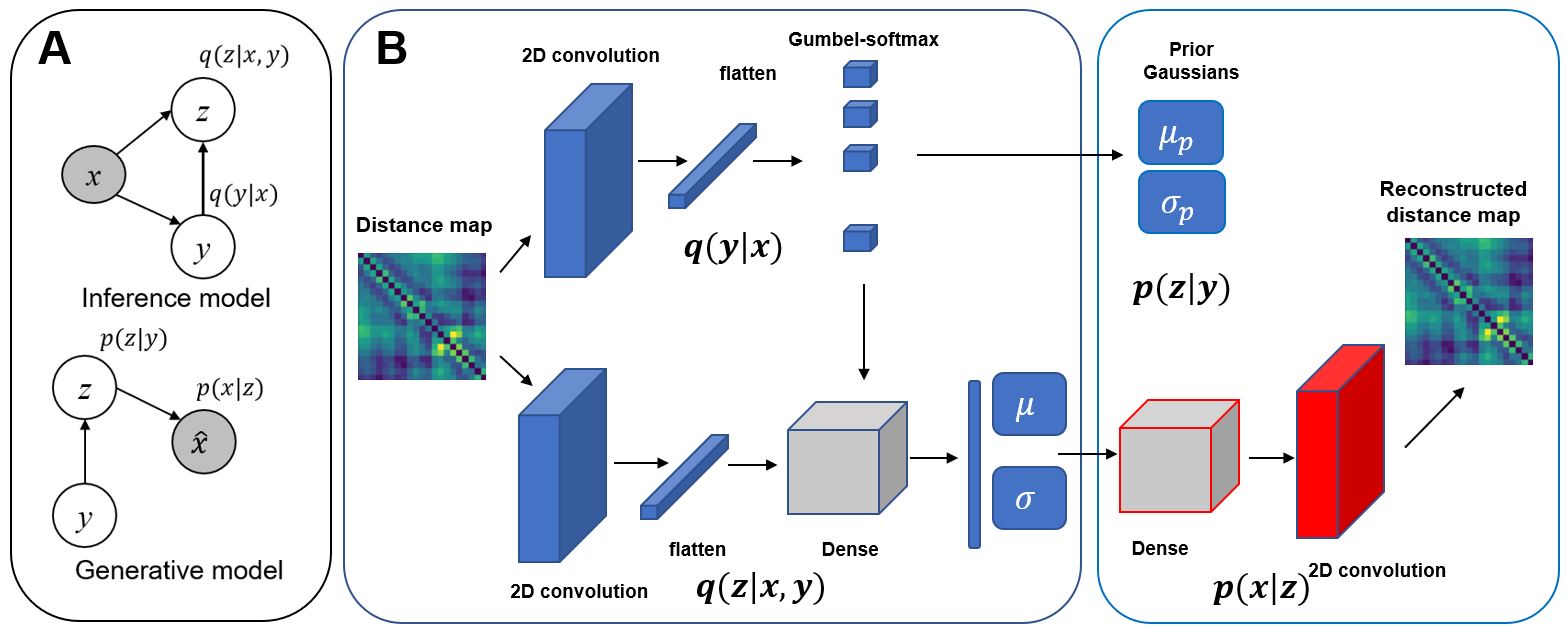}
    \caption{\textbf{A)} graphical model for inference and generative parts of GMVAE. Grey circles represents the observed data \textbf{B)} Schematic of GMVAE architecture}
    \label{fig:my_label}
\end{figure*}

\begin{table*}
\caption{best hyperparameters for each protein}
\label{tab:my-table}
\resizebox{\textwidth}{!}{%
\begin{tabular}{|c|ccccccccc|}
\hline
systems &
  \multicolumn{1}{c|}{\begin{tabular}[c]{@{}c@{}}number\\ of\\ layers\end{tabular}} &
  \multicolumn{1}{c|}{\begin{tabular}[c]{@{}c@{}}number\\ of\\ neurons\end{tabular}} &
  \multicolumn{1}{c|}{\begin{tabular}[c]{@{}c@{}}latent\\ dimension\end{tabular}} &
  \multicolumn{1}{c|}{batch-size} &
  \multicolumn{1}{c|}{temperature} &
  \multicolumn{1}{c|}{\begin{tabular}[c]{@{}c@{}}kernel\\ size\end{tabular}} &
  \multicolumn{1}{c|}{\begin{tabular}[c]{@{}c@{}}learning\\ rate\end{tabular}} &
  \multicolumn{1}{c|}{\begin{tabular}[c]{@{}c@{}}number\\ of\\ filters\end{tabular}} &
  \begin{tabular}[c]{@{}c@{}}pooling\\ sizes\end{tabular} \\ \hline
Trp-cage & 2 & 64 & 3 & 5000 & 0.1  & {[}3,3{]}   & 0.001 & {[}64,64{]}    & {[}1,1{]}   \\ \cline{1-1}
BBA      & 2 & 64 & 2 & 5000 & 0.1  & {[}3,3{]}   & 0.001 & {[}64,64{]}    & {[}2,2{]}   \\ \cline{1-1}
villin   & 3 & 64 & 3 & 2500 & 0.05 & {[}3,3,3{]} & 0.001 & {[}64,64,32{]} & {[}2,2,1{]} \\ \hline
\end{tabular}%
}
\end{table*}

\section{Results}
Here we tested the performance of GMVAE model for dimensionality reduction and clustering of three protein folding systems including Trp-Cage (pdb: 2JOF)\cite{barua2008trp}, BBA (pdb: 1FME)\cite{sarisky2001betabetaalpha} and Villin (pdb: 2F4K)\cite{kubelka2006sub}. The native folded structure of these proteins are shown in figure 2. We show that the GMVAE embedding captures the free energy landscape of these proteins with well-separated clusters. We analyze the structural properties of each cluster and show that each cluster corresponds to a different structural feature in the protein. During training, we split the data into train/validation set with a fraction of 0.8 for training and 0.2 for validation set. The total loss, cross-entropy loss and reconstruction loss shows a decreasing behavior for both train and validation set in all three proteins. We use a 2D or 3D embedding for all proteins and show that this embedding mimics the funnel-shaped landscape of protein folding where the folded state resides down in the funnel and the unfolded states are outside the funnel. GMVAE gives the class assignments probabilities for each data point. The cluster assignment is done by a K-nearest neighbors method where for each datapoint we look at the 500 nearest neighbors and calculate the sum of class assignment probabilities and finding the cluster that has the maximum probability in its neighborhood. To test whether the GMVAE clusters give meaningful structural information, we sampled 5000 datapoints from the center of each cluster and compared the distribution of RMSDs of specific parts of the protein to the native state. Moreover, we show that building a Markov model on the embedding of GMVAE produces folding and unfolding timescales that are in close agreement with the timescales obtained from constructing a Markov model on a dynamical embedding such as TICA.
\begin{figure}[h]
    \centering
    \includegraphics[scale=0.4]{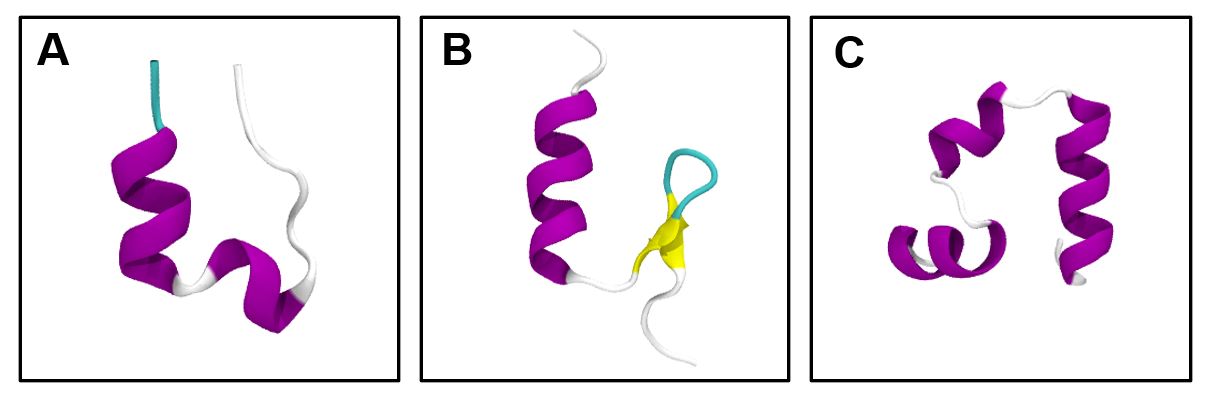}
    \caption{Native folded structure of studied proteins. \textbf{A)} Trp-cage, \textbf{B)} BBA \textbf{C)} Villin headpiece}
    \label{fig:my_label}
\end{figure}

\subsection{Trp-cage}

\begin{figure*}
    \centering
    \includegraphics[scale=0.6]{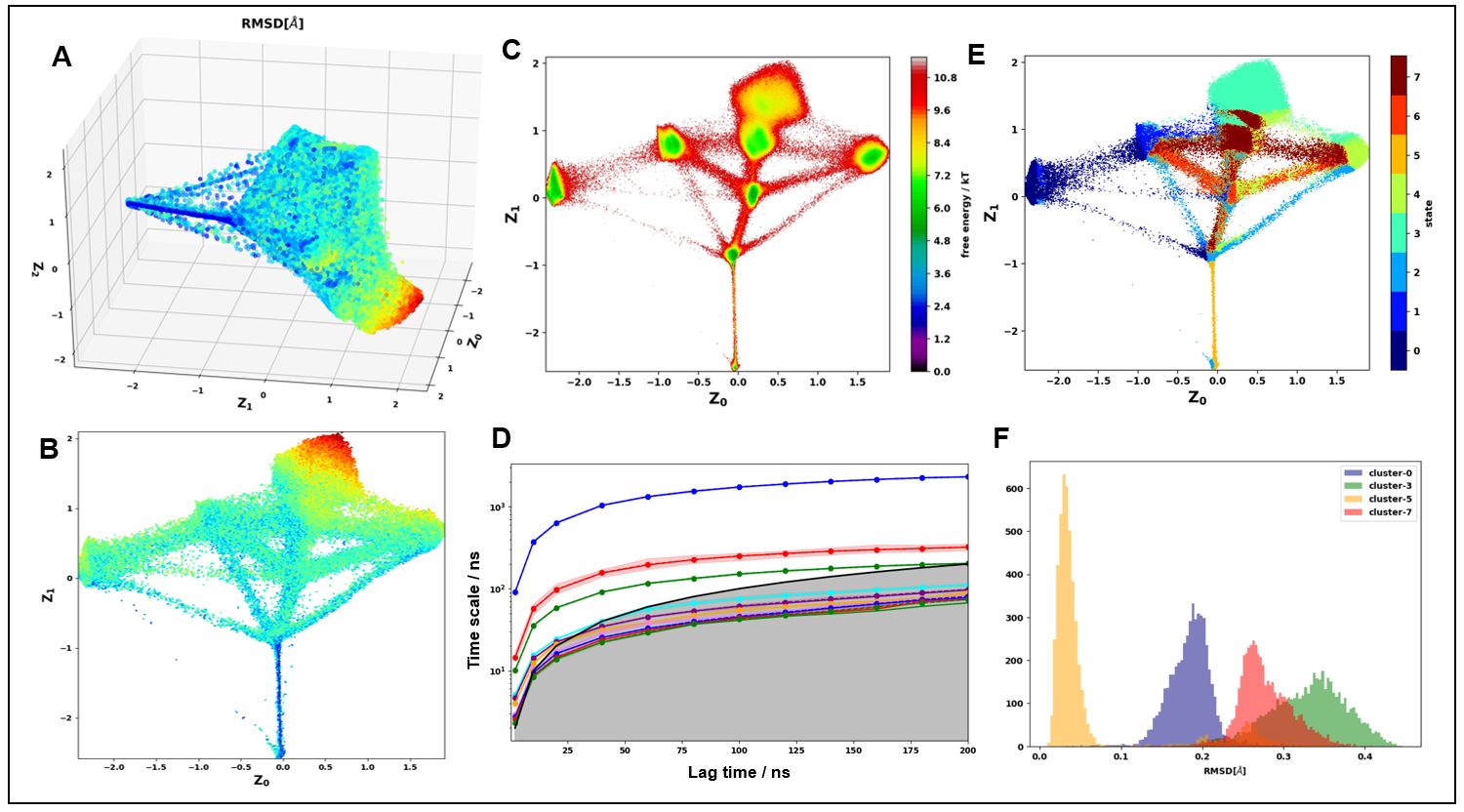}
    \caption{ \label{fig:trp_cage} Results of GMVAE for Trp-cage. \textbf{A)} 3D embedding colored with RMSD with respect to the folded state. \textbf{B)} first two dimensions of latent space colored with RMSD \textbf{C)} Free energy landscape of the first two dimensions of embedding \textbf{D)} cluster assignments based on kNN on cluster assignment probabilities \textbf{E)} RMSD distribution of $3_{10}$ residues in selected clusters \textbf{F)} implied timescale (ITS) plot for MSM construction}
\end{figure*}

As the first example, we test our GMVAE model on an ultra-long 208 $\mu s$ explicit solvent simulation of the K8A mutation of 20-residue Trp-cage TC10b at 290K by D.E. Shaw Research.\cite{lindorff2011fast} Numerous experimental and computational studies have been performed on Trp-cage.\cite{meuzelaar2013folding, english2015charged, sidky2019high} The folded state of Trp-cage shown in figure 2A contains an $\alpha$-helix (residues 2-8), a $3_{10}$ helix and a polyproline II helix, and the tryptophan residue is caged at the center of the protein. Two different folding mechanisms has been identified for Trp-cage to date \cite{deng2013kinetics}: one where Trp-cage goes through a hydrophobic collapse into a molten globule followed by formation of N-terminal helix and the native core (nucleation-condensation) and second the pre-formation of the helix from the extend conformation and the joint formation of $3_{10}$-helix and hydrophobic core (diffusion-collision). The second mechanism is identified as the dominant folding pathway for Trp-cage.

Here we investigated Trp-cage folding trajectories using GMVAE model for embedding and clustering. The features are the normalized distances between the $C_{\alpha}$ atoms of Trp-cage in the trajectories. The hyperparameter K which identifies the number of clusters is unknown \textit{a priori} and it is expected that the ensemble have a hierarchical structure depending on the level of resolution.\cite{varolgunecs2020interpretable} We tested different hyperparameters of the GMVAE model including number of clusters, batch-size, learning-rate, number-of-layers, temperature in Gumbel-softmax, kernel sizes, number-of-filters and pooling sizes for each protein. The best hyperparameters for each protein folding are selected based on the cross-entropy and reconstruction loss and also by visually looking at the embedding. For better visual inspection of the embedding we used a latent embedding dimension of 3 ($z_{dim}=3$). We picked a value of 8 for the number of clusters in this protein. The embeddings for Trp-Cage with $K=10$ and $K=7$ clusters are shown in figures S4 and S5 respectively. The total, reconstruction and cross-entropy loss using the determined hyperparameters in table 1 are shown in figure S1. Reconstruction and cross-entropy loss for both training and validation data show a decreasing behavior demonstrating the convergence of the model after 100 epochs of training.   

\begin{figure}
    \centering
    \includegraphics[scale=0.38]{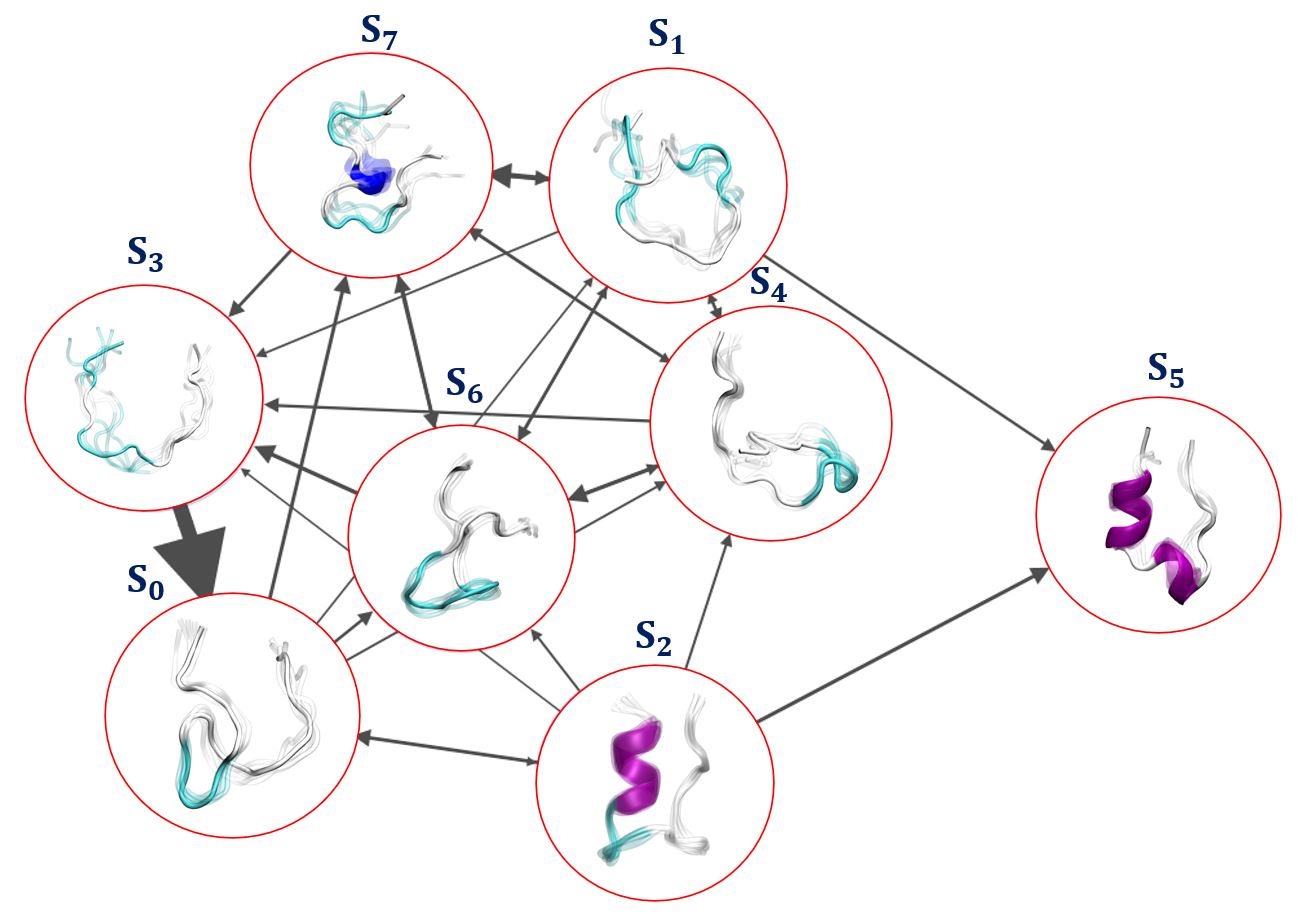}
    \caption{\label{fig:trp_cage_transitions}Trp-cage folding transitions, the thickness of lines corresponds to the transition probability between the two states. Transitions with probabilities less than 0.05 are not shown for clarity}
    
\end{figure}

Figure 3A shows the 3d embedding of Trp-cage trajectories colored based on the RMSD with respect to the crystal structure. The gradual change of color from high RMSD (red) to low RMSD (blue) in the landscape demonstrates that the low-dimensional embedding can capture the protein folding process. Additionally the shape of the low-dimensional embedding resembles the protein folding funnel where the folded state resides down the funnel and the ensemble of unfolded states are outer in the funnel. This shape can also be explained by the configurational entropy. The folded state has a narrow width in the model and the unfolded state presents a larger variance in the model due to its high fluctuation and high configurational entropy. Figure 3B shows the first two dimensions of the latent embedding colored based on RMSD. The high RMSD and low RMSD regions are well separated on this landscape. The folded state has a narrow distribution and is the narrow wedge of the folding funnel. We computed the free energy landscape on the first two dimensions of the latent space (figure 3C). The free energy landscape shows multiple wells that are separated by diffuse regions in between them. The wells correspond to the centers of GMVAE clusters and the diffuse region is the transition region between different conformational states. Since the GMVAE model gives the cluster assignment probabilities, which is the probability of each datapoint belonging to each of the K GMVAE cluster, we used a K-nearest neighbors algorithm for a hard cluster assignment. The cluster assignments is determined by computing the most probable assignment of the k neighboring datapoints. We used 500 neighbors in the k-nearest neighbors algorithm for Trp-cage embedding. The resulting clusters are shown in figure 3E. The first 2 dimensions of the embedding shows highly non-overlapping clusters.  The unfolded cluster ( cluster 3) is the extended structure and cluster 5 corresponds to the folded conformation of Trp-cage. To ensure that the GMVAE clusters correspond to different structural features of Trp cage we computed the RMSD of residues 11-15 comprising the $3_{10}$-helix for different states. The results are shown in figure 3F. State 5 has a sharp distribution corresponding to the folded conformation of these residues. Other distributions are wider and correspond to different unfolded conformations. Cluster 3 has the highest and widest distribution of RMSD and this cluster corresponds to the extended conformation of Trp-cage.

Next, we built a MSM on the 3d embedding by choosing 300 KMeans points and discretizing the trajectories based on this clustering on the GMVAE embedding. The implied timescales for this transition matrix is shown in figure 3D.  Based on this, we chose a lag time of 100 ns to build the MSM. To compute the mean-first-passage-time (MFPT) between different GMVAE clusters, we coarse grained the transition 300-state matrix into 8 states that corresponded to the GMVAE clusters. The folding and unfolding times based on the coarse-grained Markov model are 11.18 $\mu s$ and 4.30 $\mu s$. The folding and unfolding times are in good agreement with the values reported by Lindorff-Larsen et. al\cite{lindorff2011fast} who reported 14.4 and 3.1 $\mu s$ as the folding and unfolding times of this protein using the average lifetime in the folded and unfolded states observed in trajectories using a contact based definition of folded and unfolded states. A visualization of the 8 metastable states found by GMVAE model is shown in figure \ref{fig:trp_cage_transitions}. The arrows between different states show the transition between different conformations and the arrow thickness relates to the transition probability between different clusters obtained by coarse graining the Markov model into 8 GMVAE clusters. The native folded state $S_{5}$ accounts for about 18\% of the total distribution and the unfolded ensemble represents the remaining 82\%. The model predicts two folding pathways, either through a molten globule state ($S_1$) or through a pre-folded helix structure ($S_2$) which is consistent with other experimental and simulation studies. \cite{sidky2019high, marinelli2009kinetic, dickson2013native}

\subsection{BBA}
The second example is $\beta\beta\alpha$ fold protein (BBA) which is a 28-residue fast folding protein. The NMR structure of this protein is shown in figure 2B. This protein contains an antiparallel $\beta$ sheet at the N terminal and a helical conformation at its C terminus. 
Several models and hyperparameters were tested for this protein. We chose a value of 10 for the number of clusters in BBA. 2D and 3D embeddings were constructed and evaluated for this protein. This embedding resembles the folding funnel of protein with unfolded states possessing a larger distribution outside the funnel and the folded and near folded states down in the funnel. The folded ensemble is connected to multiple states which reveals multiple folding pathways of BBA and multiple intermediate states. Figure 5A shows a 2D embedding of BBA colored based on RMSD with respect to the folded structure. Unfolded and folded states are well separated on this 2D embedding. To separate different clusters in the embedding we performed a k-nearest neighbors as described earlier on the embedding. The result is shown in figure 5B which exhibits highly separated and non-overlapping clusters in the 2D embedding. In this embedding, state 4 corresponds to the folded state and state 1 is the near-folded (misfolded) state and all the other states are the unstructured or unfolded conformations. The highly non-overlapping clusters in the GMVAE landscape showcases the ability of this model to separate a vastly diverse set of protein conformations from a protein folding trajectory. The free energy landscape on this embedding is shown in figure 5C. It is observed that all clusters reside in the wells of the free energy landscape. There are also some diffuse and high energy regions between these wells which are the transitions between different metastable states. These regions are also where the model is least certain about the cluster assignment. Figure 5E shows the distribution of RMSD of antiparallel $\beta$-sheet (residues 7 to 14) (top figure) and the $\alpha$-helical (bottom figure) parts of BBA (residues 16 to 26) with respect of the folded structure. It is observed that cluster 1 has a similar distribution to cluster 4 which is the folded state. This means that cluster 1 and 4 have similar secondary structures but the orientations of the two are different. 3D embeddings with $K=9$ and $K=8$ clusters were trained for BBA and are shown in figures S6 and S7 respectively. These embeddings also resemble the folding funnel of protein folding and show highly non-overlapping clusters.

\begin{figure*}
    \centering
    \includegraphics[scale=0.6]{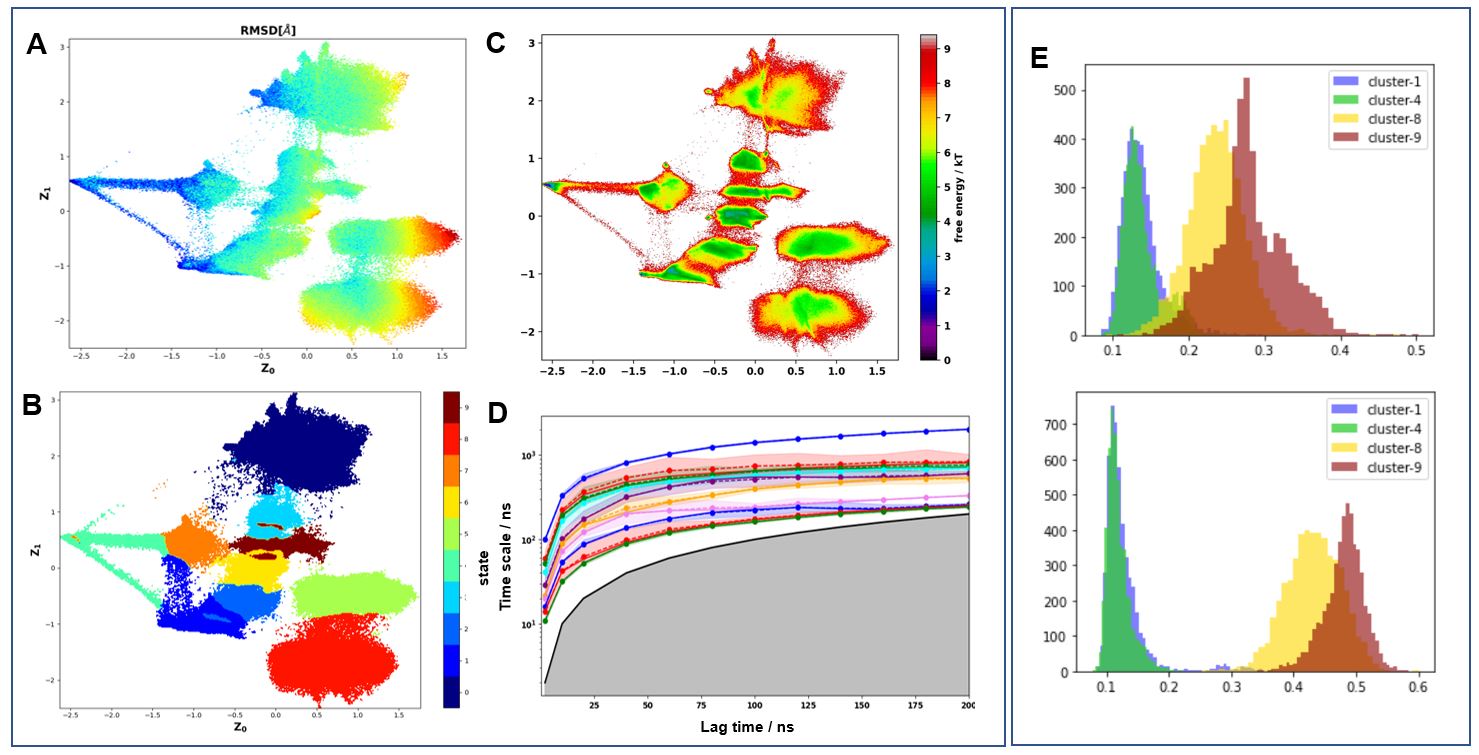}
    \caption{\textbf{A)} 2D embedding of BBA colored based on RMSD to the folded state \textbf{B)} clusters in 2D embedding of BBA using kNN for cluster assignment \textbf{C)} 2D free energy landscape of BBA based on 2D embedding \textbf{D)} ITS plot for BBA based on 2D embedding \textbf{E)} histograms for different clusters (top figure shows the RMSD for residues 7-14 or the antiparallel $\beta$-sheet and the bottom figure shows the RMSD distribution for the $\alpha$-helical residues (16-26)}
    \label{fig:my_label}
\end{figure*}

To perform a Markov model on this embedding, we first clustered this embedding using 500 KMeans and discretized the trajectories based on the points. To choose the proper lag-time for the MSM model, we plotted the implied timescales (figure 5D) and picked 200 ns and built the transition probability matrix. Next to compute the transition timescale between different GMVAE clusters, we assigned each of the 500 KMeans clusters to the closest cluster in GMVAE and then computed the mean first passage times (MFPTs) between clusters. The folding and unfolding timescales calculated here are 18.17 $\mu s$ and 8.83 $\mu s$ which are in close agreement with the values reported by DE Shaw group.\cite{lindorff2011fast} 

Figure 6 illustrates the representative structures of each cluster which are sampled from the mean of each distribution in the latent space. The transition between different states is shown with the arrow where the width of each arrow represents the transition probability. The near folded state $S_1$ has a high transition probability to the folded state $S_4$. Unfolding of BBA proceeds through transition to either $S_7$ or $S_2$ states.

\begin{figure}[h]
    \centering
    \includegraphics[scale=0.55]{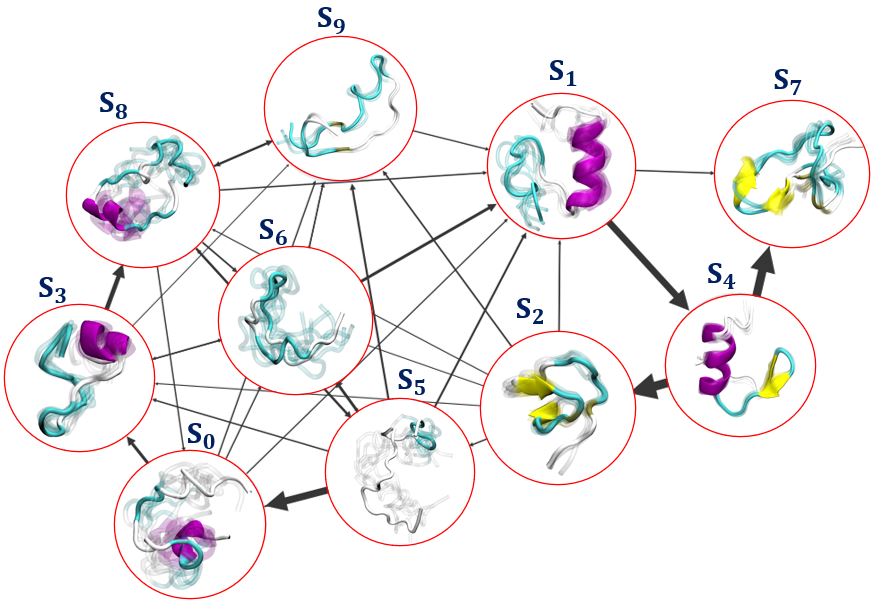}
    \caption{BBA transitions, the arrows show the transition between different clusters and the arrow thickness represents the transition probability between the corresponding clusters. Transition with probabilities less than 0.06 are not shown for clarity. Transitions with probabilities less than 0.1 are not shown for clarity.}
    \label{fig:my_label}
\end{figure}

\subsection{Villin}
The last example is a 35-residue villin-headpiece subdomain, which is one of the smallest proteins that can fold autonomously. It is composed of three $\alpha-$helices denoted as helix 1 (residues 4-8), helix 2 (residues 15-18) and helix 3 (residues 23-32) and a compact hydrophobic core. The observed experimental folding timescale for wild-type villin is about 4 $\mu s$ and the replacement of two lysine residues (Lys65 and Lys70) with uncharged Norleucine (Nle) yield a mutant with folding time of less than one microsecond. \cite{kubelka2003experimental} Folding landscape of villin double mutant has been studied both by experiments and computer simulations.\cite{sormani2019explicit,lei2010folding,chong2018examining,beauchamp2011quantitative} Folding a double mutant of Villin was studied using long-timescale molecular dynamics by D.E. Shaw group and is used here.\cite{lindorff2011fast} 

\begin{figure*}
    \centering
    \includegraphics[scale=0.6]{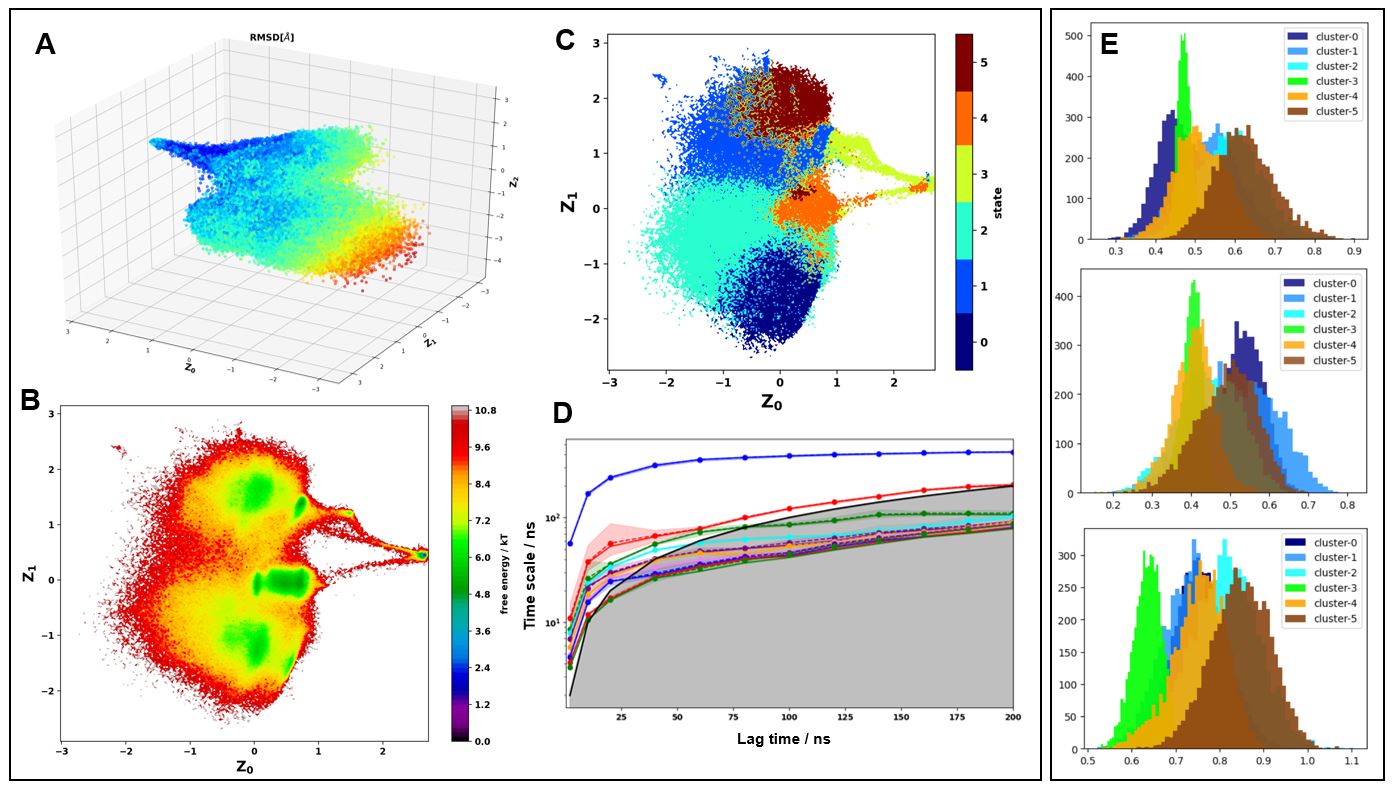}
    \caption{GMVAE embedding results for villin. \textbf{A)} 3D latent space colored with RMSD \textbf{B)} FEL based on first 2 dimensions of latent space \textbf{C)} cluster assignment using KNN \textbf{D)} ITS plot for markov model \textbf{E)} distribution of RMSD of helices 1 to 3 from top to bottom of the panel }
    \label{fig:my_label}
\end{figure*}

\begin{figure}[h]
    \centering
    \includegraphics[scale=0.55]{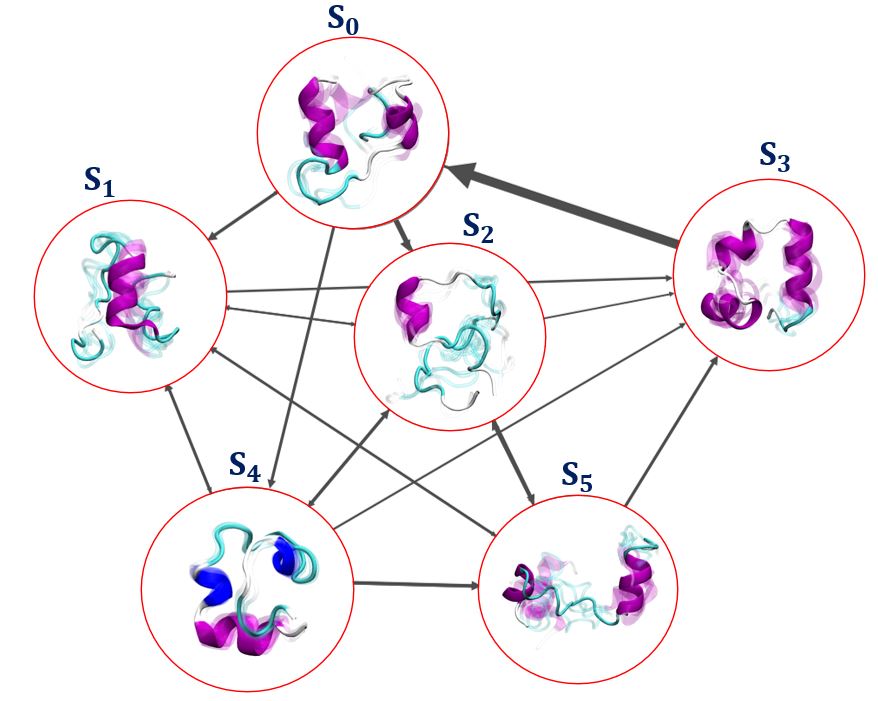}
    \caption{Transitions between different states in villin headpiece simulation. The thickness of arrows corresponds to the transition probability between the two states.}
    \label{fig:my_label}
\end{figure}
Hyperparameters for this protein were selected based on the validation loss and the cross-entropy loss and are shown in table 1. A value of 6 was chosen for the number of clusters. The training and validation loss for this protein are shown in figure S3. Latent embeddings with other hyperparameters specifically $K=8$ and $K=7$ clusters are shown in figures S8 and S9 respectively. The 3D embedding obtained from the GMVAE model is shown in figure 7A where each point was painted based on RMSD with respect to the crystal structure. Like other proteins, the 3D landscape resembles a folding funnel with folded state down the funnel and the unfolded states having larger distributions (larger conformational entropy) upper in the folding funnel. Figure 7B shows the free energy landscape on the first two dimensions of the embedding. Due to fast transitions between different states in villin, unlike BBA the FEL has larger diffuse regions with smaller basins at the center of each cluster. Presence of large diffuse regions on this landscape means that the metastable states in the folding of villin are short lived and transition between each other quickly. The K-nearest neighbors described earlier was applied to the villin embedding for a hard-cluster assignment using 500 nearest neighbors. Figure 7C shows the first two embeddings colored based on this cluster assignment. As seen, cluster 3 corresponds to the folded state and clusters 0,1,2,4,5 correspond to misfolded or unfolded ensemble. Figure 7E shows the structural properties of each cluster. specifically from the top to the bottom of this panel each distribution is the RMSD of the helix 1, 2, 3 residues with respect to the crystal structure. We observe that each cluster has a different distribution for the helical residues of the protein which are gaussian. Cluster $S_0$ has a low RMSD for helix 1 but higher RMSD values for helix 2 and 3. Secondary structure calculations showed that $S_0$ has a folded helix 1 (66$\pm$27$\%$) but unfolded helix 2 and a near-folded helix 3 ($38\pm16\%$). Most clusters have a folded or near-folded helix-3, except for cluster $S_1$ which has only about $15\%$ folded structure for helix-3. Cluster $S_3$ is the folded state where are helices are folded with more than $80\%$ probability. Moreover, $S_5$ has an unfolded helix-1 with less than $10\%$ helical probability, whereas helix-1 is folded or near-folded (more than $40\%$) in other clusters. Helix-2 is only folded in $S_3$ which shows folding of this helix is crucial for proper folding to native state. 

We next built a Markov model on this embedding by choosing 500 K-means cluster points for discretizing the trajectories. The implied timescales for this discretization are shown in figure 7D. A lag-time of 180 ns was chosen to build the transition matrix. The 500 K-means clusters were then assigned to their nearest GMVAE clusters to build a coarse-grained transition matrix. The folding and unfolding times obtained based on the constructed MSM on this embedding are 2.25 $\mu s$ and 1.54 $\mu s$ , respectively, which are in good agreement with the values reported DE Shaw group (2.8 $\mu  s$) and others building a Markov model using TICA.\cite{lindorff2011fast, suarez2021markov, pan2016demonstrating}  Figure 8 shows the structures of each cluster and the transition probability between different states. The highest transition probability $S_3\rightarrow{}S_0$ corresponds mostly to unfolding of helix 2. Therefore, proper folding of helix 2 leads to formation of native contacts and native helices. Piano et al.\cite{piana2012protein} studied the double mutant (Nle/Nle) of Villin and found a sparsely populated intermediate that involved formation of helix 3 and the turn between helices 2 and 3. This corresponds to cluster $S_4$ in our analysis that has near-folded helix-3 ($54\%$). Mori and coworkers\cite{mori2016molecular} studied the molecular mechanis for folding of Villin and the Nle/Nle double mutatnt. They found that the mutation Lys$\rightarrow{}$Nle speed up the folding transition by rigidifying helix-3. In our analysis, we found that helix-3 is folded in most clusters.

\subsection{Discussion and Conclusion}
Here we demonstrated the use of a deep learning algorithm, Gaussian mixture variational autoencoder (GMVAE), to help analyze and interpret the highly complex landscape of protein folding trajectories. The GMVAE model acknowledges the multi-basin nature of protein folding by enforcing a mixture of multiple Gaussian as the prior model for the variational autoencoder. We demonstrated our model on three long timescale protein folding trajectories, namely Trp-cage, BBA and Villin headpiece, all of which have been extensively characterized in previous studies.\cite{lindorff2011fast} In all cases, we showed that the model is able to characterize different features of the structure that could correspond to folded, misfolded or unfolded states. The low dimensional embedding obtained by GMVAE for these proteins resembles the folding funnel where the folded states lay down the funnel and unfolded ensemble are outside the funnel. This can be intuitively described from the conformational entropy point of view. The unfolded state has larger variations in the structure which causes the variance the gaussian learned by GMVAE to be larger than the folded cluster having a narrower distribution. This along with the continuity of the latent space makes landscape funnel-shaped. To verify that the clusters obtained by GMVAE correspond to different structural features of proteins during folding we computed the local RMSD of selected residues with respect to the folded structure. As expected the distribution of RMSD for different clusters follows a Gaussian where the folded state has the lowest and narrowest RMSD and the unfoleded (extended) structure has the highest and widest RMSD distribution. 
We used normalized distance maps as the features in our machine learning model which are practical ways to represent the simulation dataset of proteins. Other features such as contact maps can also be used as the input to the model which would give a lower resolution embedding due to the amount of information in the contact maps relative to distance maps. Specifically in our model we used convolutional operations which are known for their great ability to recognize and process image dataset. To make the model end-to-end differentiable we made use of Gumbel-softmax for sampling from a discrete distribution. The temperature parameter in Gumbel-softmax was tuned along with other model hyperparameters during training. The best hyperparameters for each protein was chosen based on reconstruction and cross-entropy losses. The number of clusters is also a hyperparameter and depending on the level of resolution different values can be chosen. However, one needs to take extra care when attempting different values of this hyperparameter as it can lead to high variance in the folding and unfolding times obtained by making a Markov model on the embedding. In this work, we used 2 or 3 dimensional embedding for better visualization of the latent space however, larger embedding dimensions can also be used. We tested the models with higher values of the embedding dimensions (e.g. 10), and the reconstruction loss was only slightly better than a lower dimensional embedding such as 3.  

Beyond the static characterization of the protein folding trajectories, we tested whether the model is able to characterize the kinetics of protein folding. We built a high resolution Markov model on the embedding obtained by GMVAE and computed the MFPTs between different states. Interestingly the folding timescales obtained by the model are in good agreement with the folding times reported by other groups constructing a MSM on a TICA landscape which characterizes the dynamics of folding. We should note that our model does not utilize any lag-time for construction of the low-dim embedding however, it is able to describe the folding timescales with reasonable accuracy. However, for some of the most dynamic proteins such as villin with fast folding timescales, only the first two implied timescales converge after 180 ns and the other implied timescales are below the maximum likelihood threshold which makes the model unable to give meaningful information about these faster processes. This might be remedied by adding dynamical information to the model by using a lag time in the training process. Further improvements to the model could include graph embedding of protein structures instead of using a distance map. This will be studied in a future work.

\subsection*{Acknowledgements}
This work was partially supported by the National Heart, Lung and Blood institute at the National Institute of Health for B.R.B. and M.G.; in addition, it was partially supported by the National Science Foundation (grant number CHE-2029900) to J.B.K. The authors acknowledge the biowulf high-performance computation center at National Institutes of Health for providing the time and resources for this project. We also would like to thank DE. Shaw research group for providing the simulation trajectories.

\subsection*{Conflicts of Interest}
The authors declare that there is no conflict of interest regarding the publication of this article.

\subsection*{Data Availability}
The source code for the analysis can be found at github page:
github.com/ghorbanimahdi73


\section*{References}
\bibliography{mybib.bib}
\end{document}



\title{ Supplementary Information for Variational embedding of protein folding simulations using gaussian mixture variational autoencoders}

\author{Mahdi Ghorbani}
\email{ghorbani.mahdi73@gmail.com.}
\affiliation{ 
Laboratory of Computational Biology, National, Heart, Lung and Blood Institute, National Institutes of Health, Bethesda, Maryland 20824, USA.
}%
 \affiliation{Department of Chemical and Biomolecular Engineering, University of Maryland, College Park, MD 20742, USA}
 
\author{Samarjeet Prasad}%
\affiliation{ 
Laboratory of Computational Biology, National, Heart, Lung and Blood Institute, National Institutes of Health, Bethesda, Maryland 20824, USA.
}%

\author{Jeffery B. Klauda}
\affiliation{%
Department of Chemical and Biomolecular Engineering, University of Maryland, College Park, MD 20742, USA}%
\author{Bernard R. Brooks}%
\affiliation{ 
Laboratory of Computational Biology, National, Heart, Lung and Blood Institute, National Institutes of Health, Bethesda, Maryland 20824, USA.
}%

\date{\today}

\maketitle

\maketitle
\begin{figure}[h]
    \centering
    \includegraphics[scale=0.5]{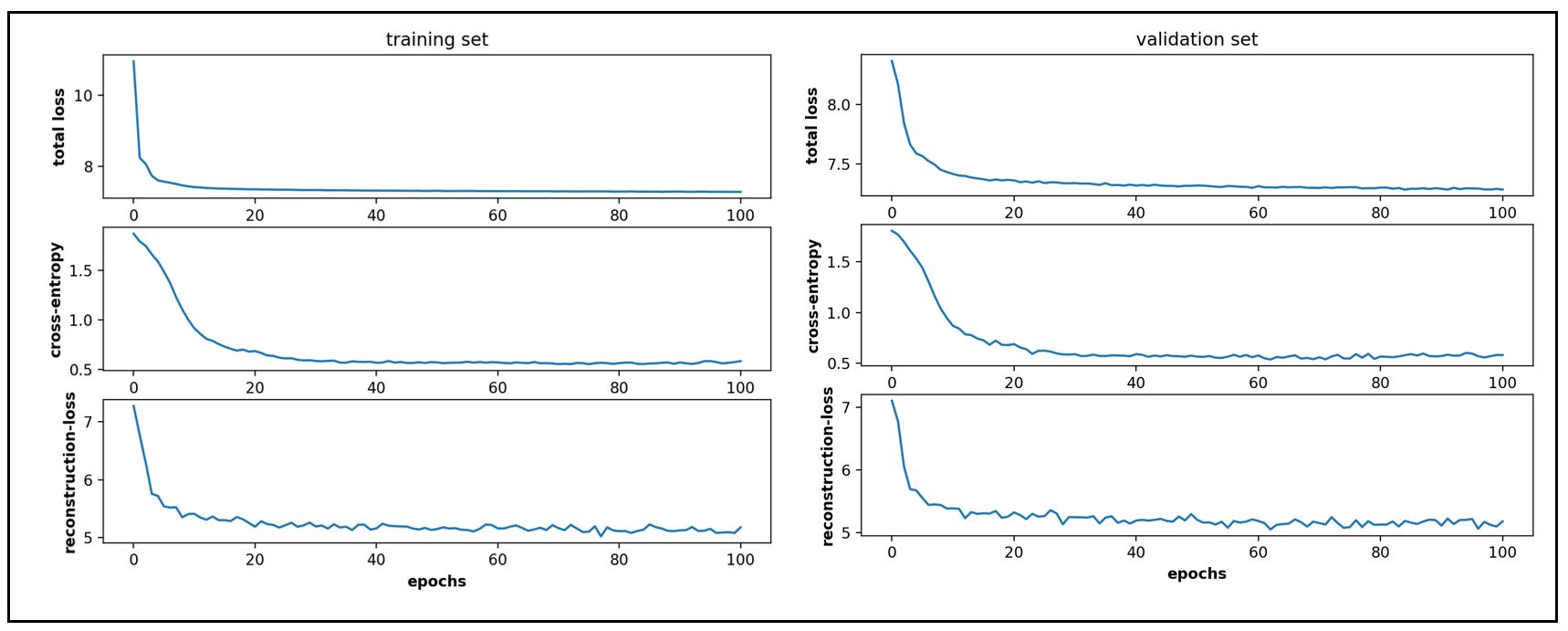}
    \caption{training and validation losses for Trp-cage}
    \label{fig:my_label}
\end{figure}

\begin{figure}[h]
    \centering
    \includegraphics[scale=0.5]{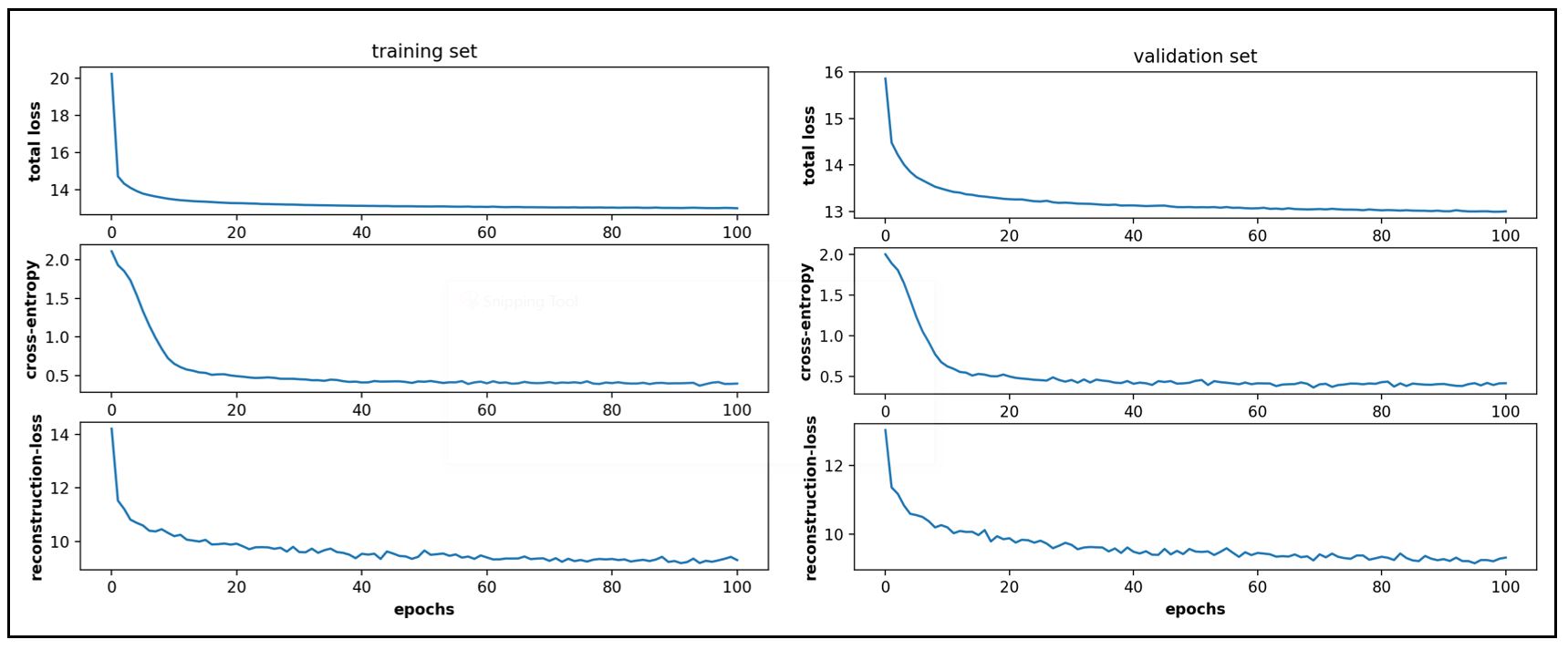}
    \caption{training and validation losses for BBA}
    \label{fig:my_label}
\end{figure}

\begin{figure}[h]
    \centering
    \includegraphics[scale=0.5]{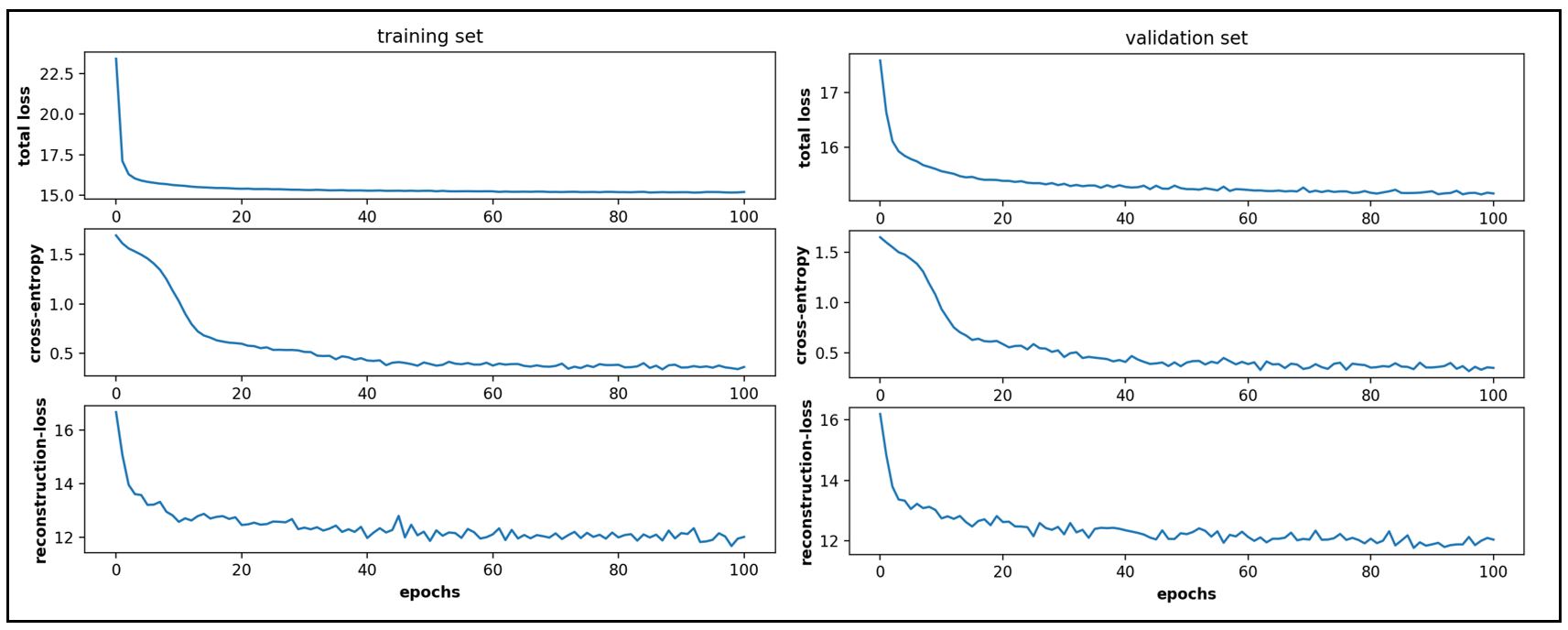}
    \caption{training and validation losses for Villin}
    \label{fig:my_label}
\end{figure}

\begin{figure}[h]
    \centering
    \includegraphics[scale=0.55]{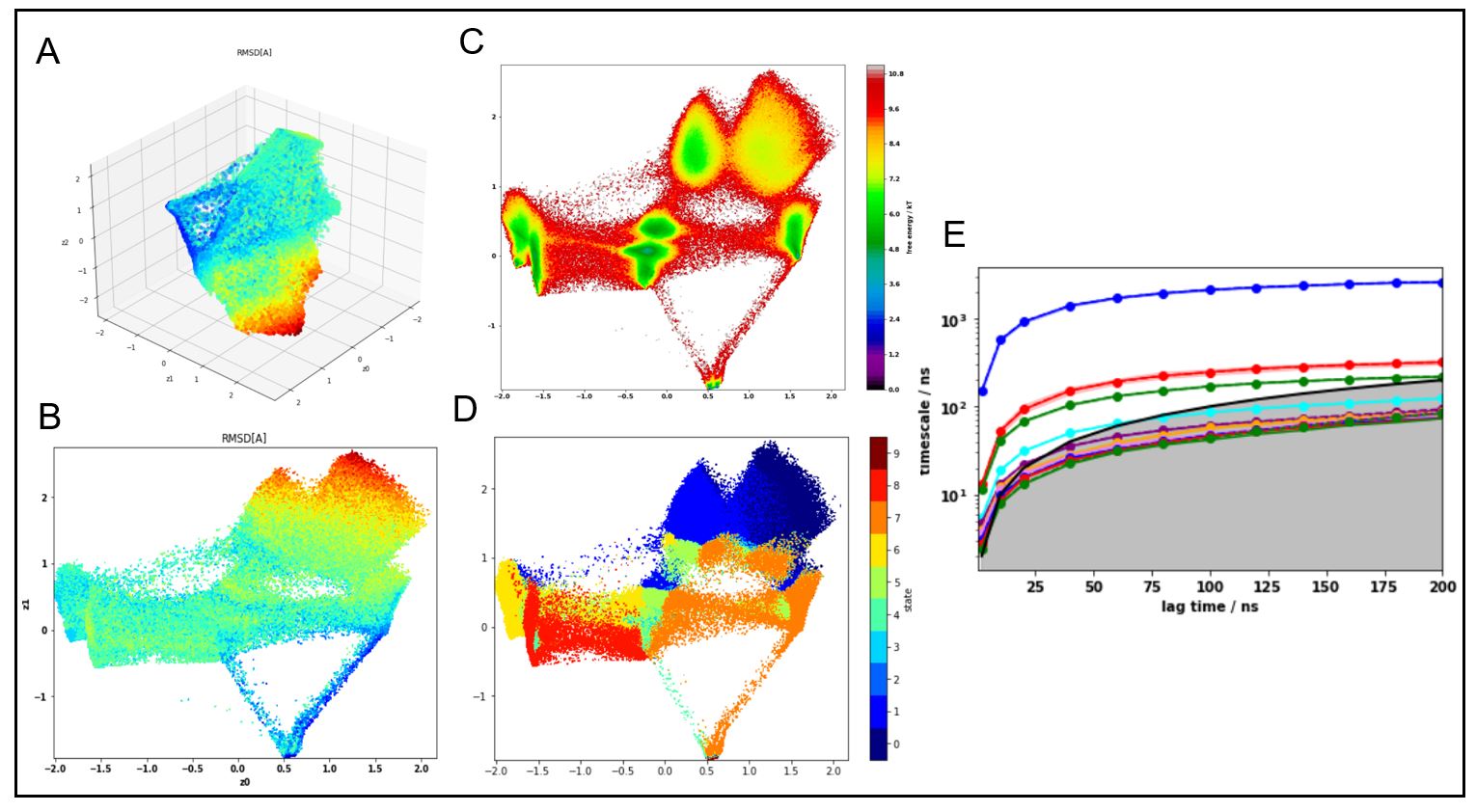}
    \caption{Trp cage embedding results from GMVAE with $K=10$ clusters and 3D embedding $z=3$ A) 3D embedding colored with RMSD B) First two dimensions of latent space colored with RMSD C) Free energy landscape on the first two embeddings D) GMVAE clusters using 500 K-nearest neighbors for cluster assignment E) Implied timescale plot for the Markov state built on the embedding using 500 KMeans points for clustering}
    \label{fig:my_label}
\end{figure}

\begin{figure}[h]
    \centering
    \includegraphics[scale=0.55]{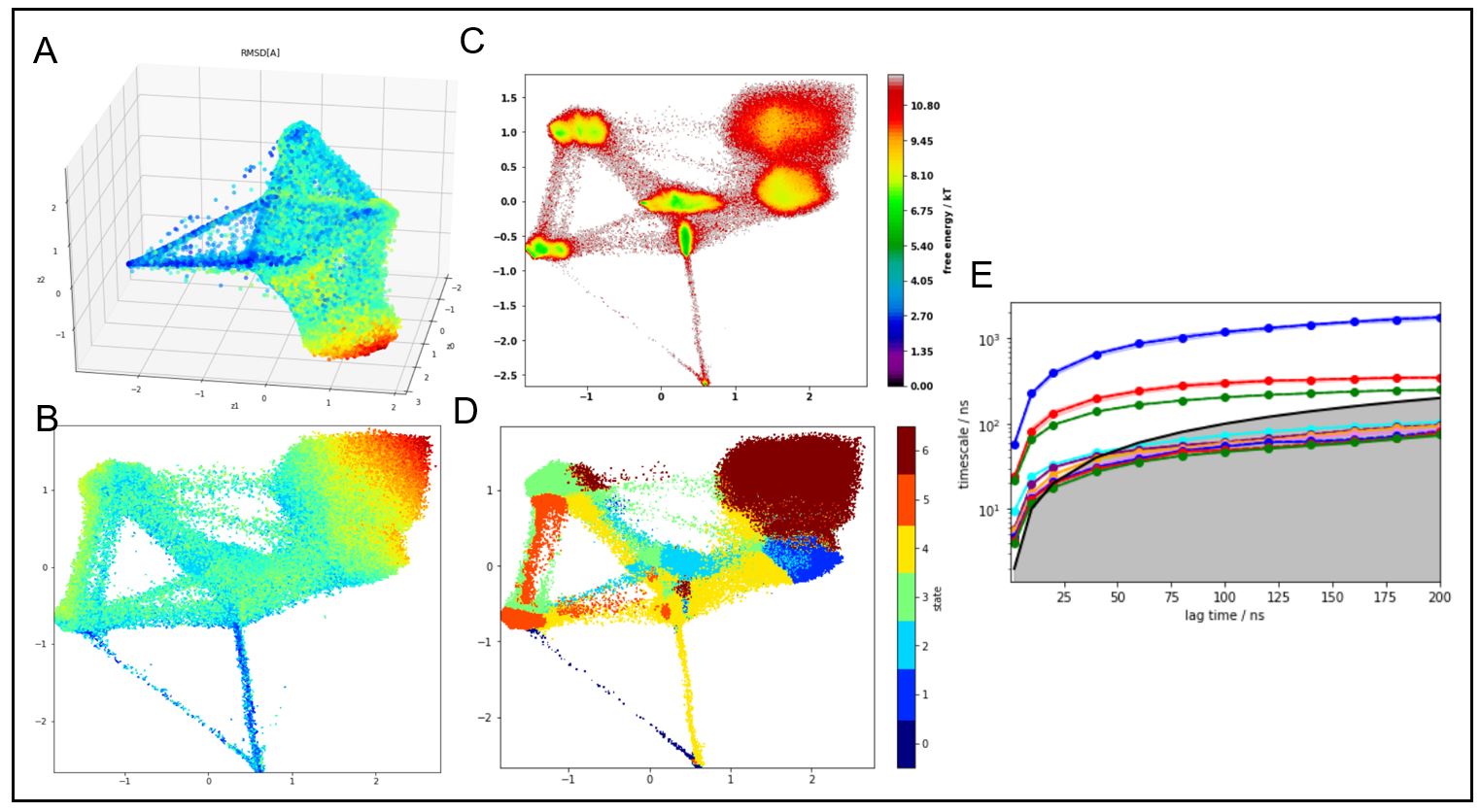}
    \caption{Trp cage embedding results from GMVAE with $K=7$ clusters and 3D embedding $z=3$ A) 3D embedding colored with RMSD B) First two dimensions of latent space colored with RMSD C) Free energy landscape on the first two embeddings D) GMVAE clusters using 500 K-nearest neighbors for cluster assignment E) Implied timescale plot for the Markov state built on the embedding using 500 KMeans points for clustering}
    \label{fig:my_label}
\end{figure}

\begin{figure}[h]
    \centering
    \includegraphics[scale=0.55]{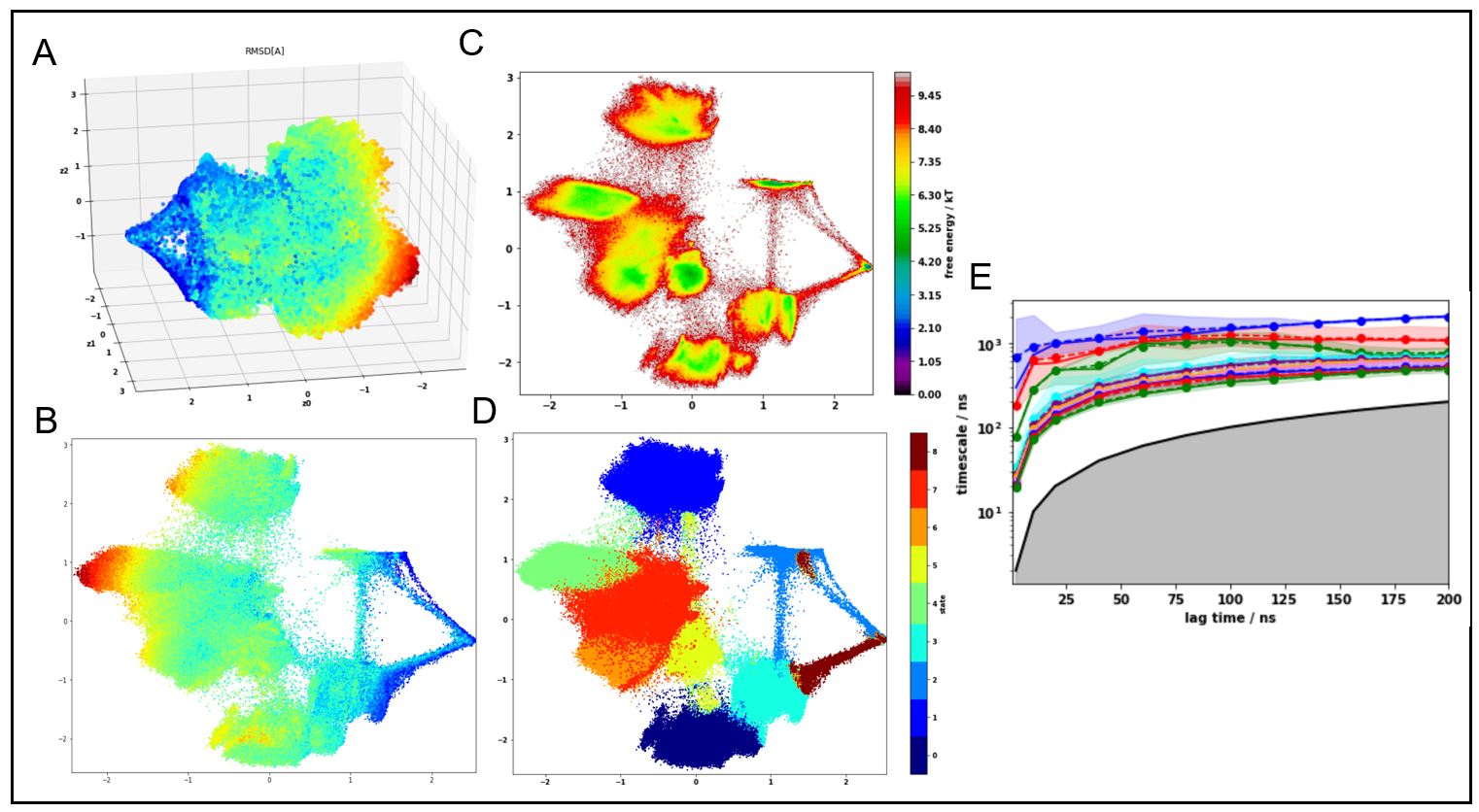}
    \caption{{BBA cage embedding results from GMVAE with $K=9$ clusters and 3D embedding $z=3$ A) 3D embedding colored with RMSD B) First two dimensions of latent space colored with RMSD C) Free energy landscape on the first two embeddings D) GMVAE clusters using 500 K-nearest neighbors for cluster assignment E) Implied timescale plot for the Markov state built on the embedding using 500 KMeans points for clustering}}
    \label{fig:my_label}
\end{figure}

\begin{figure}[h]
    \centering
    \includegraphics[scale=0.55]{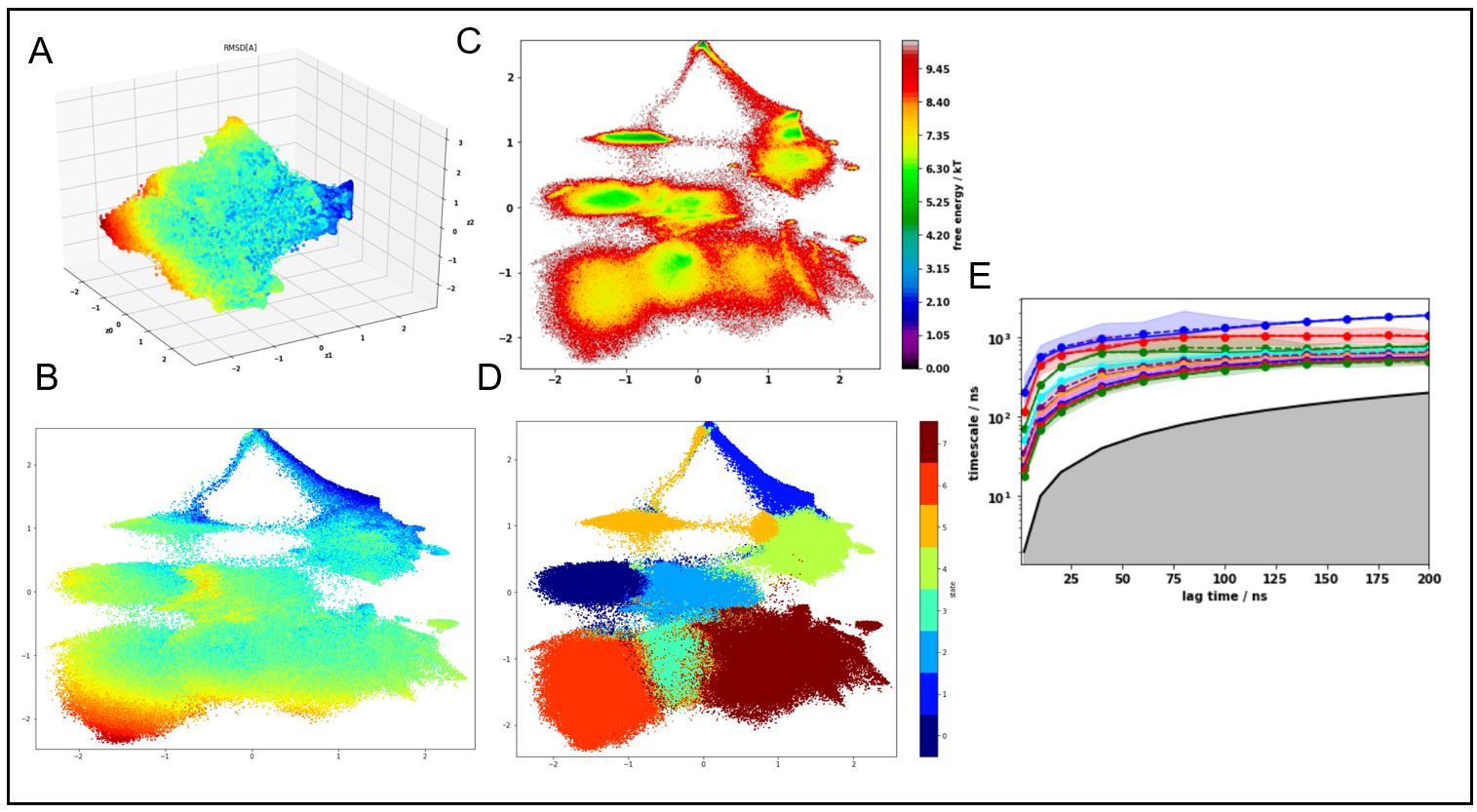}
    \caption{{BBA embedding results from GMVAE with $K=8$ clusters and 3D embedding $z=3$ A) 3D embedding colored with RMSD B) First two dimensions of latent space colored with RMSD C) Free energy landscape on the first two embeddings D) GMVAE clusters using 500 K-nearest neighbors for cluster assignment E) Implied timescale plot for the Markov state built on the embedding using 500 KMeans points for clustering}}
    \label{fig:my_label}
\end{figure}

\begin{figure}[h]
    \centering
    \includegraphics[scale=0.55]{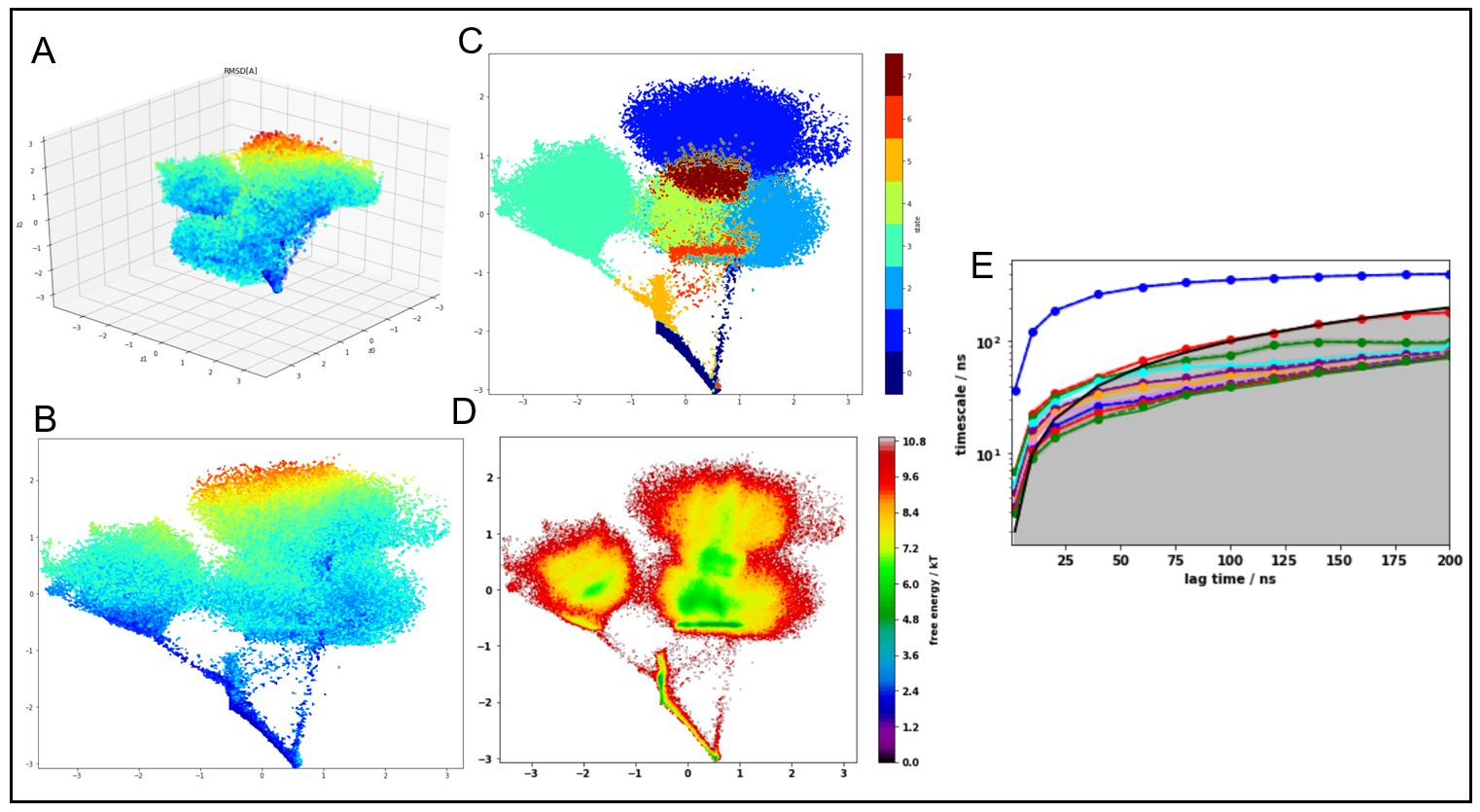}
    \caption{{Villin embedding results from GMVAE with $K=8$ clusters and 3D embedding $z=3$ A) 3D embedding colored with RMSD B) First two dimensions of latent space colored with RMSD C) Free energy landscape on the first two embeddings D) GMVAE clusters using 500 K-nearest neighbors for cluster assignment E) Implied timescale plot for the Markov state built on the embedding using 500 KMeans points for clustering}}
    \label{fig:my_label}
\end{figure}

\begin{figure}[h]
    \centering
    \includegraphics[scale=0.55]{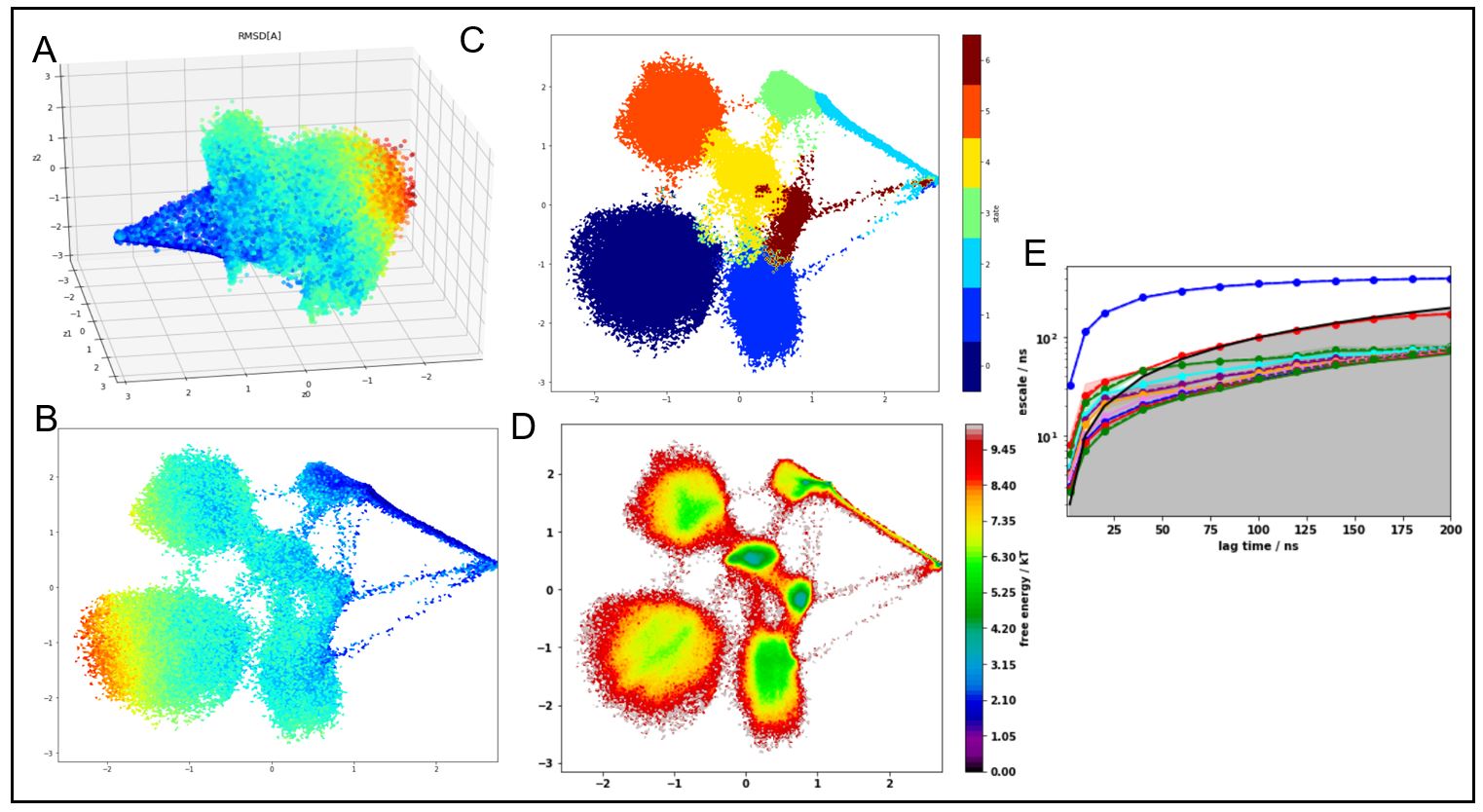}
    \caption{{ Villin embedding results from GMVAE with $K=7$ clusters and 3D embedding $z=3$ A) 3D embedding colored with RMSD B) First two dimensions of latent space colored with RMSD C) Free energy landscape on the first two embeddings D) GMVAE clusters using 500 K-nearest neighbors for cluster assignment E) Implied timescale plot for the Markov state built on the embedding using 500 KMeans points for clustering}}
    \label{fig:my_label}
\end{figure}